\documentclass{aa}  
\usepackage{hyperref}

\usepackage{graphicx}
\usepackage[export]{adjustbox}

\usepackage{txfonts}
\begin{document}

   \title{Vera C. Rubin LSST Synthetic Magnitudes derived from Gaia XP Spectra}

   \author{Oleksandra Razim
          \inst{1} \inst{2}
          \and
          Krešimir Tisanić\inst{1} \and Lovro Palaversa\inst{1}
          }

   \institute{Ruđer Bošković Institute, Bijenička cesta 54, 10000 Zagreb, Croatia\\
              \email{lp@irb.hr}
         \and
             University of Nova Gorica, Vipavska 13, SI-5000 Nova Gorica, Slovenia\\ \email{orazim@ung.si}
             }

   \date{}

  \abstract
   {In the context of Milky Way studies, the Vera C. Rubin Observatory’s Legacy Survey of Space and Time is often described as a deep extension of the Gaia survey. In the future, joint analysis of the Gaia and LSST data promises to bring new insights into our understanding of the Galaxy’s structure and formation history. However, the relatively small overlap in the magnitude ranges of the two surveys raises the question of how to perform a joint calibration of these datasets.}
   {In this paper, we produce high-quality synthetic LSST photometry from Gaia XP low-resolution spectra, using SDSS and DES observed photometry to calibrate the spectra.}
   {We develop a method of empirical correction of Gaia XP spectra using SDSS Stripe 82 photometry. We project sources for which Gaia XP data are available onto a magnitude--magnitude grid and calculate residuals between uncorrected synthetic and observed SDSS magnitudes, which are then used to derive correction coefficients across spectral regions corresponding to each photometric band.}
   {The correction significantly reduces systematic trends and scatter in the synthetic magnitudes: the median residuals decrease by an order of magnitude (e.g., for the $u$ band the improvement is from 0.038 to 0.002~mag), and the standard deviation of residuals typically becomes up to factor of two smaller (e.g., for the $u$ band from 0.2 to 0.07~mag).}
   {Our correction approach improves the reliability of synthetic photometry derived from Gaia XP spectra and enables joint analysis of Gaia and LSST data for studies of Galactic structure and stellar populations.}


   \maketitle

\section{Introduction} \label{sec:intro}

In the context of Galactic science, the upcoming Legacy Survey of Space and time (LSST) can be considered a deep counterpart to the highly successful Gaia mission, which recently completed its data-gathering programme. With its large étendue, LSST will sweep most of the southern sky in six \textit{ugrizy} bands roughly once every three nights, obtaining precise photometry for sources brighter than \textit{r} $\sim 24.5$, generating real-time alerts for transient events, and collecting data for the final 10-year catalogue, which will describe 17 billion stars and 20 billion galaxies to a $5\sigma$ depth of $\textit{r}\sim27$\footnote{See the LSST System Science Requirements Document, available at https://ls.st/srd}. While Gaia’s astrometric precision and accuracy are unattainable by the ground-based surveys, LSST's photometric depth, combined with repeated observations in six photometric bands, enables a multitude of distance-measurement methods applicable all the way to the edge of the stellar halo. These methods include those based on standard candles, such as RR Lyrae, Cepheids, and long-period variables \citep[e.g.][]{Rau2019,Musella2024}, as well as methods based on photometric parallaxes \citep[e.g.][]{2008ApJ...673..864J,Bailer-Jones2021}, including a fully Bayesian method for the estimation of distance, metallicity, and interstellar dust extinction along the line of sight developed with LSST-sized datasets in mind \citep{2025AJ....169..119P}. Combined with maps of stellar densities and metallicities of main-sequence stars out to distances of up to 100 kpc, and with high-quality observations of large samples of intrinsically faint objects (e.g. brown dwarfs, low-mass stars, and white dwarfs), LSST data promise to deliver a wealth of information about the structure and history of the Milky Way and stellar evolution processes, continuing the advances in this area initiated by the Gaia mission.

In this context, a joint analysis of these two surveys is of great interest; however, they have a relatively small overlap in their magnitude ranges: Gaia’s faint limit lies around $r\sim20$, while the LSST saturation limit is close to $r\sim16$. Only $\sim1\%$ of the sources observed by LSST will be brighter than $r\sim20$. Considering also that the two surveys use different photometric systems, combining them is not a trivial task.

One way to address this is to use the low-resolution spectra provided with Gaia Data Release 3 \citep{DeAngeli2022,Montegriffo2022}. This dataset contains information for about 220 million sources, and these spectra can be used not only to derive a wide range of astrophysical parameters, but also to produce synthetic photometry in any photometric system within the optical wavelength range\footnote{Excluding photometric systems with bandpass widths narrower than the line spread function of the XP spectra.}. By generating synthetic LSST magnitudes from these low-resolution spectra, we create a catalogue suitable for direct comparison and calibration of Gaia and LSST observations.

Although conceptually simple, delivering a well-calibrated synthetic photometry is fraught with pitfalls. One possible approach, successfully implemented on Gaia data by \cite{Montegriffo2023} through \texttt{GaiaXPy}, is based on the concept of standardisation \citep{Bessell2005}. 
Alternatively, adjusting the spectra instead of the transmission curves should, in principle, yield equivalent results. 

We adopt this second approach and test its validity using observed SDSS and DES photometry, together with the synthetic standardised SDSS photometry produced with \texttt{GaiaXPy}. Furthermore, we use the results of our investigation to deliver Gaia XP-based all-sky LSST synthetic photometry\footnote{For the 220 million sources from Gaia DR3 with XP spectra} based on our adjusted XP spectra. The delivered catalogue not only precedes LSST Data Releases, but also provides reliable photometry in the LSST system for sources that will never be observed by the Vera C. Rubin Observatory (either because of their location or their brightness). Additionally, we anticipate that this catalogue may prove useful in LSST photometric calibration.

The paper is organised as follows. In Section \ref{Data}, we describe the Gaia XP, SDSS Stripe 82, and DES datasets used to produce and correct synthetic photometry, together with the quality cuts applied. In Section \ref{Production}, we present the calibration of the Gaia XP spectra and the details of our correction procedure. In Section \ref{Results}, we compare uncorrected and corrected synthetic magnitudes with observed ones, estimate median residuals, and investigate remaining magnitude- and colour-dependent biases. We summarise the results in Section \ref{Conclusions}. Additional technical details on data processing are provided in Appendix \ref{appendixFilters}, and additional plots illustrating the quality of various derived catalogues are placed in Appendix \ref{appendixPlots}.

\section{Data}\label{Data}

\subsection{Gaia XP spectra}\label{gaiaxpspec}

As part of Data Release 3, Gaia delivered approximately 219 million mean low-resolution spectra acquired by the BP and RP spectrophotometers onboard the spacecraft \citep{DeAngeli2022}. Most of the spectra correspond to sources brighter than 17.6 mag in the Gaia $G$ band, except for a small sample of ultra-cool dwarfs with magnitudes as faint as $G=21.4$. The BP spectrophotometer covers the wavelength range $\lambda=[330,680]$ nm with a resolving power $(\lambda/\Delta\lambda)$ ranging from about 100 at the blue end to $\approx25$ at the red end. For the RP spectrophotometer, the wavelength range is $\lambda=[640,1050]$ nm, and the resolving power $(\lambda/\Delta\lambda)$ varies from about 100 at the blue end to $\approx65$ at the red end. The signal-to-noise ratio of the time-averaged mean spectra varies significantly with wavelength, brightness, and colour of the observed sources. Generally, the highest-quality spectra are those in the $9 < G < 12$ magnitude range \citep{DeAngeli2022}.

At the internal calibration stage, performed by the Gaia team, each mean spectrum is brought onto a common flux and pseudo-wavelength scale. Internally calibrated spectra are stored as linear combinations of basis functions (the so-called continuous representation) and can be accessed via the Gaia Archive\footnote{\url{https://gea.esac.esa.int/archive/}}. The Python package \texttt{GaiaXPy}\footnote{\url{https://gaia-dpci.github.io/GaiaXPy-website/}, \url{https://gaiaxpy.readthedocs.io/en/latest/}} \citep{gaiaxpy,Montegriffo2023}, developed to facilitate the use of BP/RP spectra, can be used to generate sampled versions of the spectra in internal and absolute flux and wavelength systems, and to calculate synthetic photometry in various photometric systems. The package is actively maintained, and the reader is referred to the documentation for the latest developments.

The current understanding of the instrumental model used for the processing of spectra has certain limitations and will likely improve with the processing of new Gaia observations. Both systematic biases and random flux errors in the final XP spectra have been documented. For a detailed discussion of these effects, we refer the reader to \cite{Montegriffo2022}. We point out a few issues that are particularly relevant to this work. Specifically, there is a known colour-dependent systematic at $\lambda < 400$ nm, the exact origin of which remains uncertain; a magnitude-dependent systematic at wavelengths just before the region of overlap of the BP and RP spectra ($500 < \lambda < 600$ nm), likely caused by imperfect calibration in this transition region; and a magnitude-dependent systematic at $\lambda > 950$ nm. 

Additionally, when spectra are converted to sampled form, correlations between the coefficients of the basis functions of the continuous representation are amplified. This results in high-amplitude noise in certain parts of the sampled spectra, the so-called wiggles \citep{DeAngeli2022,Montegriffo2022}. BP mean spectra, especially in the blue part, tend to exhibit more wiggles than RP spectra, although for many sources, wiggles are also noticeable at $\lambda > 1000$ nm. In some cases, these wiggles result in negative flux values in the \textit{u} and \textit{z} bands \citep{DeAngeli2022}, and their intensity does not correlate with source brightness, spectral energy distribution shape, or signal-to-noise ratio \citep{Montegriffo2022}. As such, wiggles become a significant source of scatter in synthetic magnitudes. In Sect.~\ref{synthMagsExt}, we discuss several approaches to minimising these errors.

To compensate for differences between synthetic magnitudes derived in different photometric systems, the \texttt{GaiaXPy} package provides standardised transmission curves for several large-scale surveys, such as SDSS. The essence of standardisation lies in using an external set of standard stars with precise photometric measurements to calculate residuals between observed and synthetic photometry, after which the transmission curves are adjusted to minimise these residuals (see Sect.~2.2 of \citealt{Montegriffo2023} and references therein). It is therefore expected that simply convolving XP fluxes with the published LSST filter transmission curves would lead to departures similar to those observed for non-standardised synthetic SDSS photometry\footnote{With the exception of the LSST \textit{y} band, which is not part of SDSS photometry}.

Since the first data obtained with the LSST Camera is expected in summer 2026 (so-called Data Preview 2\footnote{\url{https://rtn-011.lsst.io/}}), we adopt a different approach and adjust the spectra instead of modifying the filter throughputs, using SDSS Stripe 82 standard-star (S82) measurements as the ground truth for the correction.

\subsection{SDSS Stripe 82 standards catalogue}\label{Stripe82}

The SDSS Stripe 82 catalogue, originally presented in \cite{IvezicStripe82}, covers a relatively large region of the sky, provides a good magnitude-range overlap with the Gaia XP sample, and has small photometric errors. It contains averaged \textit{ugriz} SDSS \citep{SDSS2000} photometry for over one million non-variable stars located in the region of the sky known as Stripe 82 ($-60^{\circ} \leq \mathrm{RA} \leq 60^{\circ}$ and $-1.266^{\circ} \leq \mathrm{DEC} \leq 1.266^{\circ}$). We use a version of this catalogue presented in \cite{T21}, which is based on SDSS Data Release 15 \citep{SDSSDR15} and constructed from a factor of 2--3 more epoch observations per source compared to the original version. The catalogue was calibrated and validated against a number of external photometric datasets from various surveys, including Gaia, and was shown to have sub-percent internal photometric precision, with magnitudes extending to \textit{r}$\approx22$.

\subsection{Stellar atmosphere models}\label{Kurucz}

Gaia XP spectrophotometry covers a wavelength range of 320--1080 nm\footnote{\url{https://www.cosmos.esa.int/web/gaia/edr3-passbands}}, which is very similar to that of the LSST design bandpasses (320--1100 nm). A similar situation holds for the SDSS bandpasses, which cover a slightly broader range (300--1100 nm; \citealt{Doi2010}). The SDSS {\it z} band is effectively split into the LSST {\it z} (780--945 nm) and {\it y} (880--1098 nm) bands (according to \texttt{rubin\_sim}\footnote{\url{https://github.com/lsst/rubin_sim}}).

Because XP spectra do not cover the entire wavelength range of SDSS and LSST photometry, and because issues in XP spectra are often most pronounced at the shortest and longest wavelengths, we replace and extend the spectral tails of XP spectra with appropriate models from the Kurucz \texttt{ATLAS} catalogue of stellar atmospheres\footnote{Available at \url{https://www.stsci.edu/hst/instrumentation/reference-data-for-calibration-and-tools/astronomical-catalogs/kurucz-1993-models}} \citep{Castelli2003}. The optimal wavelength ranges for replacing observed fluxes with model fluxes are determined through a grid search. We refer to this procedure as the Kurucz extension. The \texttt{ATLAS} catalogue contains approximately 7600 stellar atmosphere models covering wavelengths from 1000~\AA\ to 10~$\mu$m, with wavelength steps varying from 2 to 5 nm. The models in the \texttt{ATLAS} catalogue are computed on a non-uniform grid of stellar parameters spanning temperatures $3500 \leq T \leq 50\,000$~K, metallicities $-5.0 \leq [M/H] \leq +1.0$, and surface gravities $0 \leq \log g \leq 5$. Early experiments showed that this grid is too coarse for our purposes; we therefore interpolate between the models with the closest stellar parameters before performing the extension.

\subsection{Sample selection and cross-match}

The sample used to derive the corrections is obtained by cross-matching high-quality SDSS Stripe 82 sources described in \cite{T21} with Gaia XP spectra. The applied quality cuts select objects with more than 15 Gaia observations, fewer than 10\% observations contaminated, high signal-to-noise SDSS Stripe 82 photometry, and a unique match between the two catalogues. Details of the cross-matching procedure, together with the ADQL queries used, are provided in Appendix~\ref{appendixFilters}. The final sample used to derive the correction coefficients contains 146\,550 objects.

\subsection{Dark Energy Survey sample}\label{DES}

As an additional validation, we use the corrected Gaia XP spectra to produce synthetic magnitudes in the Dark Energy Survey (DES; \citealt{Abbott2021}) photometric system and compare them with Data Release~2 observations. DES observes approximately 5000~deg$^{2}$ of the southern sky in the optical and near-infrared wavelength range, primarily targeting fainter objects than those in SDSS Stripe~82. The Stripe~82 region also overlaps with DES, allowing a three-way validation of our method.

We cross-matched the Stripe~82 and Gaia XP samples with the DES catalogue, selecting objects with a single best match in each survey within a 1~arcsecond search radius. Only high-quality DES measurements were used, applying the filtering criteria recommended by the DES team (\texttt{flags\_b<4} and \texttt{imaflags\_iso\_b=0}). These criteria retain close pairs and blended objects while discarding sources with saturated pixels in some of the images \citep{Abbott2021}. The final matched sample contains 80\,723 objects.

\subsection{LSST Data Preview 1 (DP1) sample}\label{LSST_DP1}
At the time of writing this article, the only photometric catalogue already released by Rubin Observatory is the Data Preview 1 (DP1) \citep{LSSTDP1}. We use it for a preliminary comparison with our corrected synthetic magnitudes. 

The DP1 catalogue was derived from science-grade exposures made with Rubin Commissioning Camera (LSSTComCam, \citet{LSSTComCam}) during a several-week observing run in 2024. It is worth noting that LSSTComCam, used in DP1 cannot be considered a miniature version of the LSSTCam, which will be used by LSST throughout its operational phase.

The DP1 Object catalogue \citep{LSSTDP1_object} contains 2\,299\,726 records. Out of these, we select only those that pass point source filters and have no flags indicating issues during shape and extendedness determination (for the full query, see Appendix~\ref{appendixFilters}). After that, we crossmatch this sample with the full Gaia DR3 sample using 1-to-1 crossmatch with only neighbours within a 1-arcsecond radius retained, which returns 8\,183 objects. After discarding objects that do not have Gaia XP spectra, only 2\,440 remain. We refer to this sample as DP1. This sample is significantly fainter than the S82 and DES samples: most of the DP1 objects have Gaia $G>16$. 

\section{Producing synthetic SDSS and LSST magnitudes}\label{Production}

\subsection{Deriving synthetic magnitudes and their quality metrics}

In the simplest case, synthetic fluxes in a set of photometric bands can be obtained from the observed spectra using the following equation:

\begin{equation}
	F_b=\int\limits_0^\infty \phi_b(\lambda)F_{\nu}(\lambda)\,\mathrm{d}\lambda 
	\label{eq:eq1}
\end{equation}

where $F_b$ is the flux in a band $b$, $\phi_b(\lambda)$ is the bandpass throughput, and $F_{\nu}(\lambda)$ is the observed flux density of a source. To compensate for the magnitude- and colour-dependent terms mentioned in Sect.~\ref{gaiaxpspec}, we determine a correction function using the observed high-precision SDSS photometry. The resulting equation for the corrected synthetic flux then becomes:

\begin{equation}
	F_b^*=\int\limits_0^\infty \phi_b(\lambda)F_{\nu}^*(\lambda)\,\mathrm{d}\lambda 
	= \int\limits_0^\infty \phi_b(\lambda)F_{\nu}(\lambda)K(\lambda)\,\mathrm{d}\lambda
\end{equation}

where $F_{\nu}^*(\lambda) = F_{\nu}(\lambda)K(\lambda)$ is the corrected flux density, obtained by multiplying the original spectrum by the correction function $K(\lambda)$. To obtain integrated fluxes, we use the \texttt{rubin\_sim} package\footnote{\url{https://github.com/lsst/rubin_sim}}\citep{rubinsim}, developed by the LSST collaboration for future LSST data processing. We apply it to spectra that are externally calibrated and converted to sampled form using \texttt{GaiaXPy}.

To determine the difference between the synthetic and observed SDSS S82 magnitudes, we define magnitude residuals ($\Delta_{\rm mag}$):
\[
\Delta_{\rm mag}=m_{\rm obs}-m_{\rm synth},
\]
where $m_{\rm obs}$ and $m_{\rm synth}$ denote the observed and synthetic magnitudes, respectively.

As quality metrics, we use two statistical indicators: the median magnitude residual ($med^{synth}_{resid} = median(\Delta_{\rm mag})$) and a robust scatter metric of magnitude residuals,
\[
\sigma = \frac{P_{84.13}-P_{15.87}}{2},
\]
where $P_{84.13}$ and $P_{15.87}$ are the percentiles of $\Delta_{\rm mag}$. We use this metric as a proxy for the standard deviation to diminish the impact of outliers and enable meaningful comparison with \cite{Montegriffo2023}. For Gaussian distributions, this metric equals the standard deviation.

In our analysis (Sect.~\ref{validation}), we find significant offsets and trends when a simple transformation of XP fluxes to SDSS magnitudes is applied (Equation~\ref{eq:eq1}). It is also worth noting that \texttt{rubin\_sim} uses the sampled form of the spectra to calculate synthetic magnitudes, while \texttt{GaiaXPy} uses the continuous form. As mentioned earlier, the sampled representation is more affected by noise, so without additional corrections, magnitudes produced by \texttt{GaiaXPy} are expected to be more accurate. In the following sections, we describe our method for correcting the XP spectra, under the assumption that the XP spectra can indeed be corrected using the S82 standard stars as a benchmark.

The starting point of the process leading to the generation of the LSST synthetic photometry catalogue is the sampled Gaia XP spectra in the absolute reference system, obtained using \texttt{GaiaXPy}. The noisy tails of these spectra are extended and partially replaced with Kurucz stellar atmosphere models, and synthetic SDSS magnitudes are calculated from these extended spectra. By comparing the synthetic magnitudes with the observed S82 magnitudes, we calculate residuals $\Delta_{\rm mag}$ and obtain a magnitude-dependent correction function for the spectra. Finally, we use it to correct the sampled spectra and calculate corrected synthetic S82 magnitudes. This correction strongly diminishes the biases, and the corrected spectra are then used to produce LSST photometry.

\subsection{Gaia XP spectra calibration and conversion to sampled form}

After Gaia collects raw BP/RP data, it first performs internal calibration to bring all spectra onto a consistent Gaia-defined system by correcting for detector sensitivity differences, wavelength shifts, and temporal variations. During the next step, called external calibration, a library of Spectrophotometric Standard Stars (SPSS) with precisely known fluxes is used to determine the absolute wavelength scale and sensitivity curve. For further details, we refer the reader to \citep{DeAngeli2022}.

We obtain the internally calibrated spectra from the Gaia Archive in continuous form, and then perform external calibration and conversion to sampled form using the \texttt{GaiaXPy} package \citep{gaiaxpy,Montegriffo2022}\footnote{\url{https://gaiaxpy.readthedocs.io/en/latest/description.html}}. Following that, during the correction procedure, we use Kurucz stellar atmosphere models to extend our spectra. These models are defined with a 2~nm step in the range 330--1000~nm and a 5~nm step in the range 1000--1050~nm, resulting in 50--60 points per SDSS band. For simplicity, we use the same sampling to transform the continuous XP spectra to the absolute system as used for the Kurucz models\footnote{We also tested a finer sampling (uniform, with a 1~nm step) and found that it does not produce any noticeable differences in the synthetic magnitudes.}.

\subsection{Producing synthetic SDSS magnitudes for extended spectra}\label{synthMagsExt}

As we explained in Sect.~\ref{Kurucz}, Gaia XP spectra do not cover the full wavelength range of SDSS or LSST photometry in the \textit{u} band. To avoid systematic underestimation of the \textit{u} magnitude, we extend the Gaia XP spectra with model SEDs from the \texttt{Kurucz ATLAS} catalogue.

First, from the Gaia Archive, we download stellar parameters for sources in our Gaia XP catalogue for which they are available ($\sim 90\%$ of all sources with Gaia XP spectra). These stellar parameters were determined by the Gaia team from BP/RP spectra using the \texttt{GSP-Phot} software package \citep{Andrae2022}. We then select the Kurucz models with the closest values for all three stellar parameters defining the models: T, $[M/H]$, and $\log g$, and interpolate between them, using the parameters from the Gaia Archive as weights. The Gaia Archive also provides the monochromatic extinction parameter $A_{0}$, determined at 541.4~nm (\texttt{azero\_gspphot}). We use it to redden the final interpolated Kurucz model, using the Fitzpatrick extinction law \citep{Fitzpatrick1999} implemented in the \texttt{extinction} Python package \citep{BarbaryExtinct}\footnote{GitHub: \url{https://github.com/kbarbary/extinction}}. We use the default ratio of total to selective extinction, $R_V = 3.1$.

In this way, we obtain a synthetic SED with a shape that should be as close to the calibrated Gaia XP spectra as possible. However, we also have to account for the distance at which the source is located, and scale the model flux values to match the observed spectrum. We do this using \texttt{rubin\_sim} functions for flux normalisation (scaling); the scaled flux is calculated as:
\[
F_{\rm scaled} = F_{\rm orig} \cdot 10^{-0.4\,(mag_{\rm obs} - mag_{\rm synth})}.
\]

If model selection and extinction were perfect, the shape of the synthetic SED and observed spectrum would be identical, and it would not matter which band is used for scaling. However, neither model selection nor extinction is perfect. In order to obtain a smooth transition between the central, observed part of the spectrum and synthetic extensions at the tails, we use two scaled versions of the model. The first is scaled using the \textit{g} band and is used to extend the blue end of the spectra, and the second is scaled using the \textit{z} band and is used to partially replace and extend the red end.

\begin{figure*}    \includegraphics[width=0.9\textwidth, center]{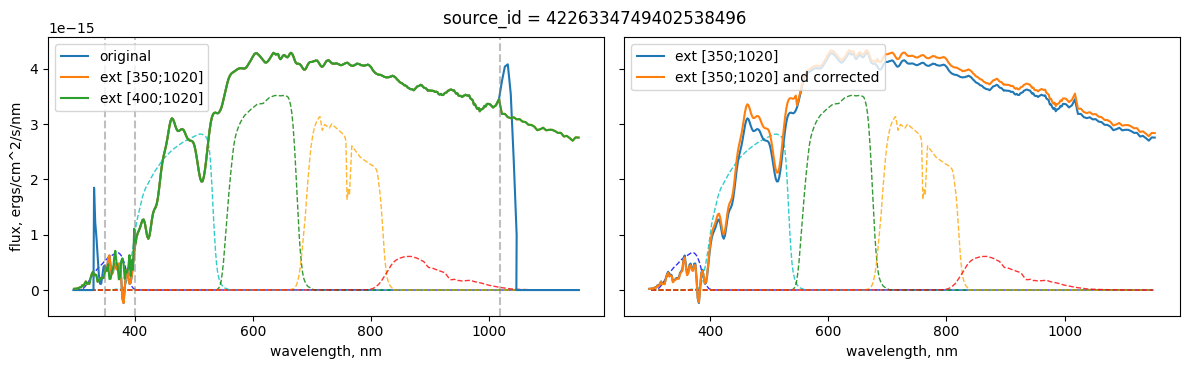}
    \caption{Left panel: an example spectrum before and after Kurucz extension, performed with different cutoff values in the \textit{u} band. Vertical dashed lines indicate the cutoff values. Right panel: the same spectrum after the [350;1020] extension, before and after the correction.}
    \label{fig:spectraTransform}
\end{figure*}

After that, one more parameter must be selected, namely the wavelength cutoff for the ``wiggly tails''. We tested multiple versions of the cutoff values, starting from no cutoff at all (i.e. synthetic extensions are simply added to the observed spectra for the part of the SDSS and LSST wavelength range not covered by Gaia XP spectra) and ending with completely replacing the \textit{u} and \textit{z} spectral regions. We found that for the \textit{u} band, replacing a larger part of the SED with a model reduces the scatter of the residuals, most likely because wiggles introduce additional noise. We obtain the best result by replacing all flux below 400~nm with model fluxes, effectively replacing the entire \textit{u}-band part of the SED. Since this may be undesirable for some applications, we also publish a catalogue in which the flux is replaced only in the $[296,350]$~nm range, as well as a catalogue with no replacement at all. The left panel of Fig.~\ref{fig:spectraTransform} demonstrates the effect that the two cutoffs have on the shape of the spectra in the \textit{u} band.

At the same time, cutting off large parts of the spectrum at the red end introduces more scatter, especially for faint sources, likely due to imperfect stellar atmosphere models or possibly a mismatch between the chosen extension and the true stellar atmosphere of the source. The residuals are minimised for the \textit{z} band with a moderate cutoff that replaces fluxes in the $[1020,1100]$~nm range.

After performing the extension, we again produce synthetic SDSS magnitudes. For the $\sim10\%$ of sources for which stellar parameters are not available, only non-extended magnitudes are calculated.

\subsection{Correcting spectra for magnitude- and colour-dependent terms using observed SDSS Stripe 82 magnitudes}\label{specCorr}

\begin{figure*}
    \includegraphics[width=0.98\textwidth, center]{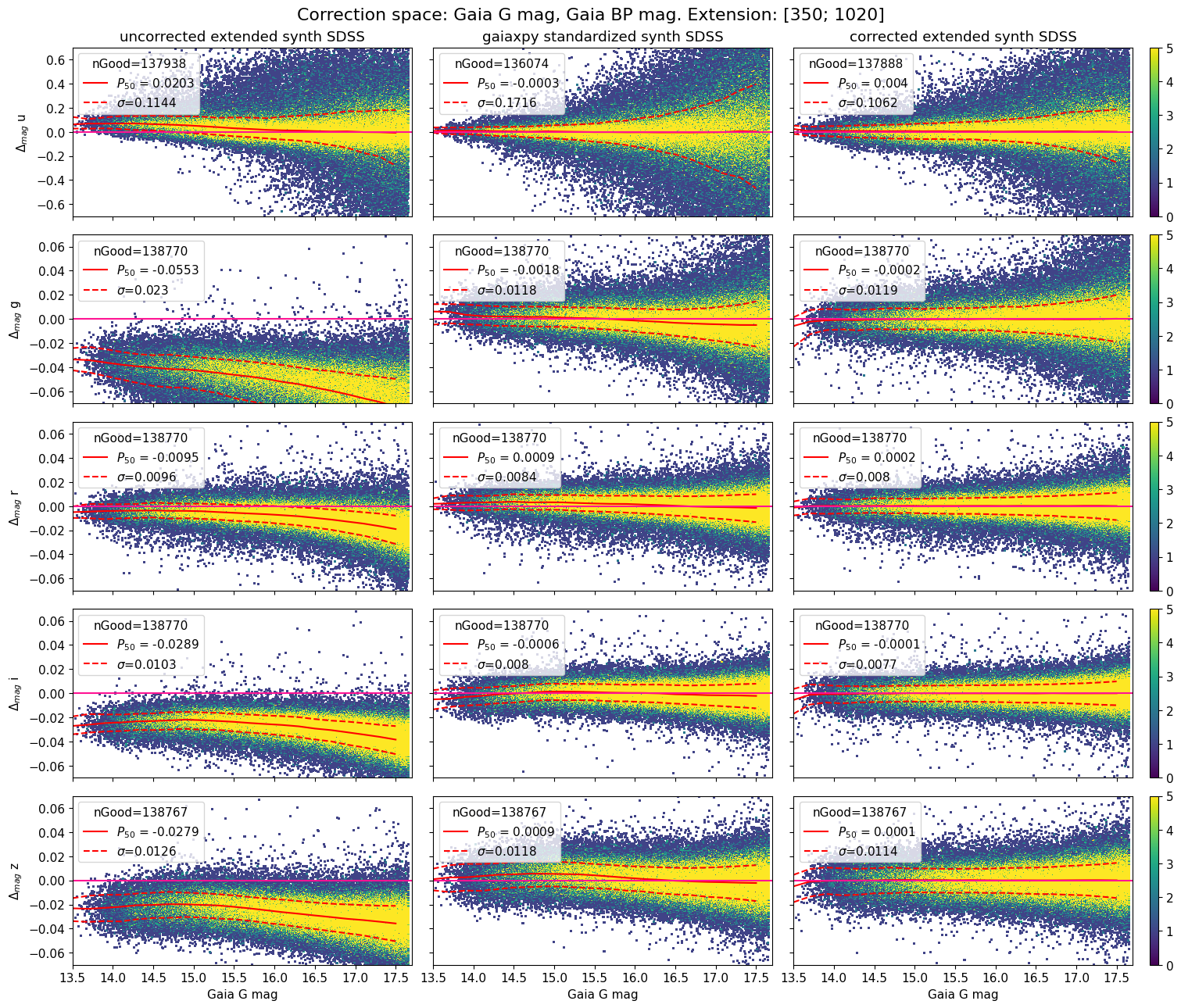}
    \caption{Magnitude-residual plots for [350;1020] extended spectra. Left column: \texttt{rubin\_sim} synthetic SDSS magnitudes with Kurucz extension but without correction. Central column: \texttt{GaiaXPy} synthetic SDSS magnitudes, using standardised filters. Right column: \texttt{rubin\_sim} synthetic SDSS magnitudes with Kurucz extension and after the correction described in Sect.~\ref{specCorr}. Note that for the \textit{u} band, the scale of the $y$-axis differs from those for the \texttt{griz} bands.}
    \label{fig:350_mag}
\end{figure*}

\begin{figure*}
    \includegraphics[width=0.98\textwidth, center]{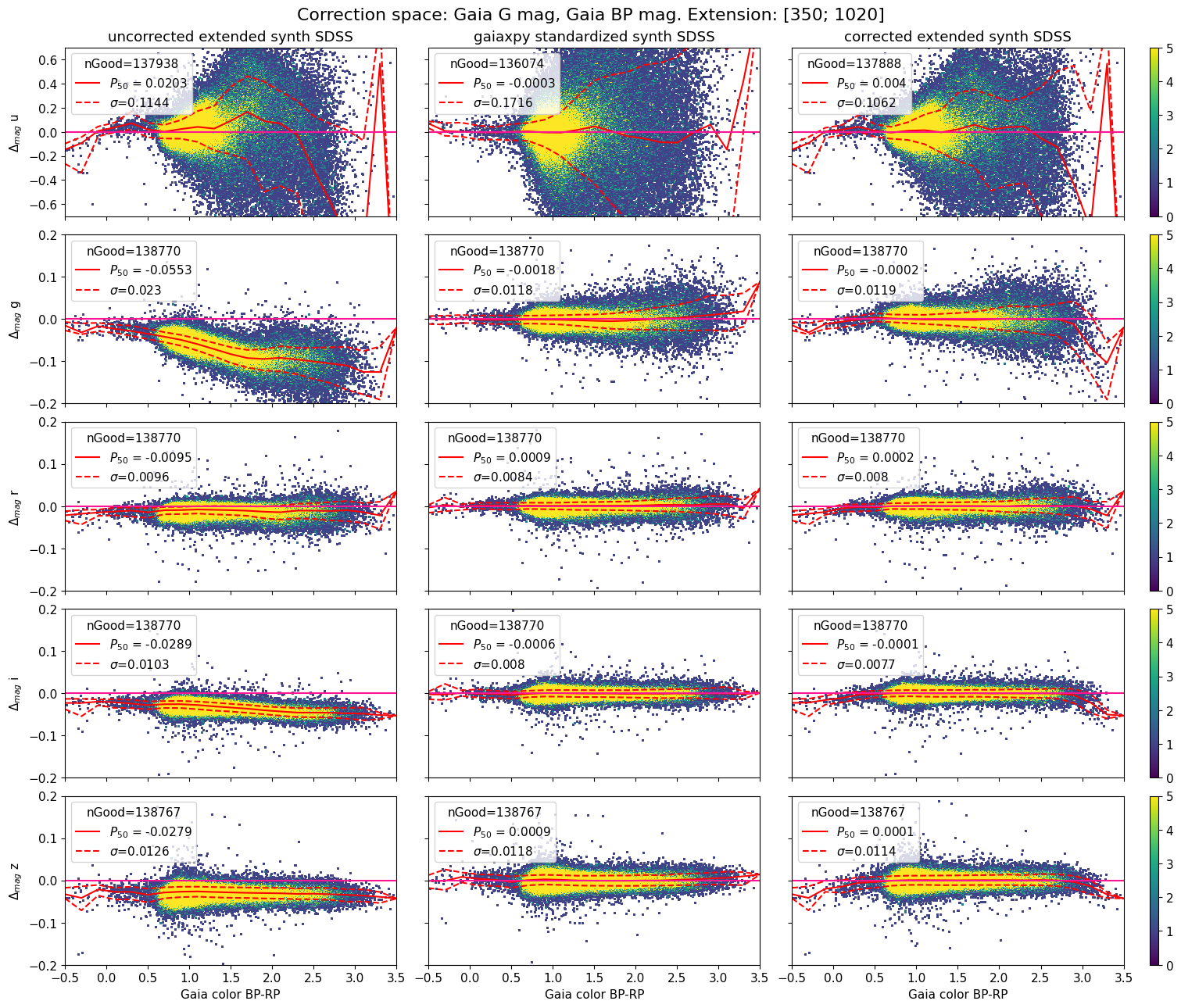}
    \caption{Colour-residual plots for [350;1020] extended spectra. Left column: \texttt{rubin\_sim} synthetic SDSS magnitudes with Kurucz extension but without correction. Central column: \texttt{GaiaXPy} synthetic SDSS magnitudes, using standardised filters. Right column: \texttt{rubin\_sim} synthetic SDSS magnitudes with Kurucz extension and after the correction described in Sect.~\ref{specCorr}. Note that for the \textit{u} band, the scale of the $y$-axis differs from those for the \texttt{griz} bands.}
    \label{fig:350_color}
\end{figure*}

To illustrate trends in the residuals of synthetic SDSS photometry produced from uncorrected XP spectra with respect to the \cite{T21} S82 standards, we apply Equation~\ref{eq:eq1} and plot residuals as a function of magnitude and BP$-$RP colour in the left columns of Figs.~\ref{fig:350_mag} and \ref{fig:350_color}. The most obvious trend is with magnitude, which is likely a result of systematic overestimation of the background that produces a negative offset in the measured XP fluxes \citep{DeAngeli2022}. Additionally, there are colour-related terms of varying strength, with the strongest effect visible in bands covering $\lambda \lessapprox 400$~nm. This is likely due to the low throughput and relatively complex transmission curve of the BP spectrophotometer in that wavelength range.

We tried several correction approaches, starting with a correction function expressed as a sum of basis functions of different types and ending with machine-learning methods. However, we obtained the best results with a numerical correction on a mag--mag grid.

In this approach, we bin sources with XP spectra on a two-dimensional grid in \textit{G} and \textit{BP} magnitude. For each bin, we calculate the correction coefficient $k_b$ for each of the five SDSS bands:
\begin{equation}
    k_b = \frac{f_{\rm obs}^{b}}{f_{\rm synth}^{b}} = 10^{-\frac{\mathrm{median}(\Delta_{\rm mag}^b)}{2.5}},
\end{equation}
where the index $b$ represents one of the SDSS \textit{ugriz} bands, and $\mathrm{median}(\Delta_{\rm mag}^b)$ is the median magnitude residual in a given bin.

To obtain a smooth correction in the \textit{G}$-$\textit{BP} space, we interpolate the values of the correction coefficients for each of the five bands when evaluating the correction for each source. We found that linear interpolation is sufficient provided that the density of the magnitude--magnitude grid is high enough. We obtain the best results with 50 cells along each axis, equivalent to a magnitude step of $\sim0.13$. Increasing the number of cells does not provide improvement and, beyond some point, starts to deteriorate the results, since too few objects fall into each cell.

The position of the object on the magnitude--magnitude diagram then determines its five correction coefficients. The resulting correction function for each spectrum is a piecewise-constant function:

\begin{align}
   K(\lambda) =
   \begin{cases}
        k_u & \text{if } 350.0 \le \lambda \le 388.2\\
        k_g & \text{if } 388.2 \le \lambda \le 545.6\\
        k_r & \text{if } 545.6 \le \lambda \le 682.0\\
        k_i & \text{if } 682.0 \le \lambda \le 826.8\\
        k_z & \text{if } 826.8 \le \lambda \le 1200.0\\
    \end{cases}
\end{align}
The right panel of Fig.~\ref{fig:spectraTransform} shows an example spectrum before and after the correction.

We experimented with different combinations of magnitude--magnitude, colour--magnitude, and colour--colour grids for this procedure; however, the differences in the results are negligible. We also implemented a three-dimensional version of this algorithm and used the three Gaia magnitudes, or combinations of two magnitudes and one colour, as axes. However, the three-dimensional version yields worse corrections than the two-dimensional one, most likely due to the increased sparsity of objects in three-dimensional parameter space.

\section{Results}\label{Results}

\subsection{Comparison with SDSS Stripe 82 data}\label{validation}
For validation of the quality of the corrected spectra, we calculate the residuals between our synthetic SDSS magnitudes and observed ones from the SDSS Stripe 82 catalogue and then compare them with the residuals between observed magnitudes and synthetic ones obtained with the standardised SDSS throughputs from the \texttt{GaiaXPy} package. Figs.~\ref{fig:350_mag} and \ref{fig:350_color} illustrate this comparison with residual vs. magnitude and residual vs. colour diagrams, respectively. The synthetic spectra used there are extended with cutoff values $[350;1020]$ nm. 

For the non-extended and uncorrected spectra (see Tab.~\ref{tab:stat_noExt}), synthetic SDSS magnitudes demonstrate significant median and scatter deviations from the observed Stripe 82 magnitudes. The scatter is one to two orders of magnitude larger for the \textit{u}-band than for other bands. In all bands, except \textit{u}-band, the residuals are negative, meaning that synthetic fluxes are underestimated, while in the \textit{u}-band the flux is slightly overestimated, especially for the bright sources.

\begin{figure*}
    \includegraphics[width=0.85\textwidth, center]{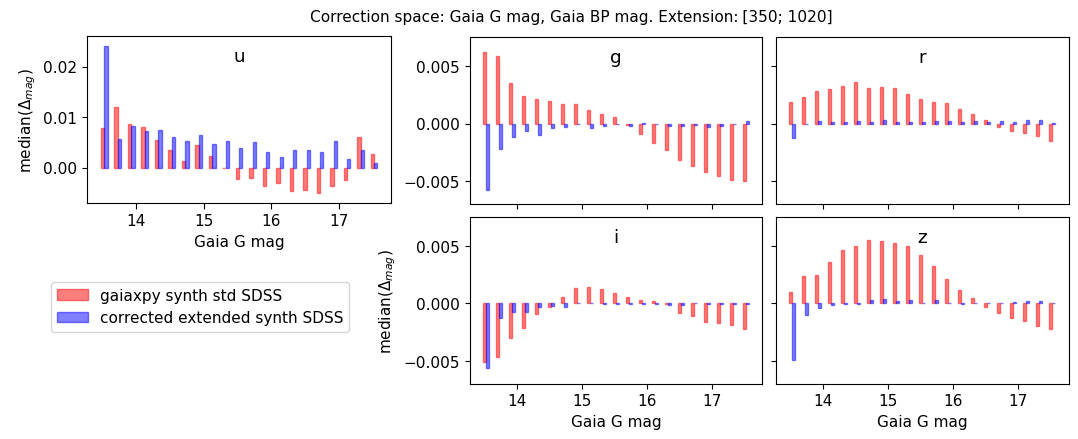}
    \caption{Median magnitude residuals in magnitude bins for synthetic SDSS photometry. Red bars correspond to the \texttt{GaiaXPy} synthetic magnitudes obtained with standardised SDSS filters, blue bars correspond to \texttt{rubin\_sim} synthetic magnitudes after the correction, described in Sect.~\ref{specCorr}.}
    \label{fig:350_bin_mag}
\end{figure*}

Tail extension with Kurucz models reduces scatter for both \textit{u} and \textit{z} bands, but especially \textit{u} (see Tab.~\ref{tab:stat_350_1020} and Tab.~\ref{tab:stat_400_1020}). Combining it with the correction described in Sect.~\ref{specCorr}, we achieve a strong reduction in the residual biases for all bands. The scatter metric drops by $\sim 20-30\%$ for \textit{riz}-bands and almost by a factor of 2 for the \textit{ug}-bands, and the median offset is improved by an order of magnitude for the \textit{u}-band and by two or three orders of magnitude for other bands. Fig. \ref{fig:350_bin_mag} shows the median residuals within individual magnitude bins. The results are comparable (for the \textit{u}-band) or better (for the rest of the bands) than those achieved with \texttt{GaiaXPy} standardised SDSS throughputs. Standard deviations within magnitude bins are very close for the two methods, apart from the faint part of the \textit{u}-band, in which our results deliver a slight improvement over \texttt{GaiaXPy}.

The Tables~\ref{tab:stat_noExt}, \ref{tab:stat_350_1020}, and \ref{tab:stat_400_1020} report the median and standard deviations of magnitude residuals in each band for each of the extension cases, before and after the correction. The best result is obtained with the [400;1020] extension. We note that the only significant difference between the three tables exists for the \textit{u}-band, and for extensions [350;1020] and [400;1020] \texttt{griz} bands are completely identical.

\begin{table}
\caption{Median and standard deviations for the $\Delta_{mag}$ without correction (\textit{synth}) and with correction (\textit{corr}). }\label{tab:stat_noExt}
\centering
\begin{tabular}{ |c||c|c|c|c|c| } 
 \hline
Band & nGood & $med^{synth}_{resid}$ & $med^{corr}_{resid}$ & $std^{synth}_{resid}$ & $std^{corr}_{resid}$ \\
 \hline\hline 
u & 143833 & 0.0388 & 0.0062 & 0.2045 & 0.1419 \\
g & 145227 & -0.0555 & -0.0002 & 0.0236 & 0.0122 \\
r & 145227 & -0.0095 & 0.0002 & 0.0099 & 0.0081 \\
i & 145227 & -0.0289 & -0.0001 & 0.0104 & 0.0078 \\
z & 145224 & -0.0327 & -0.0001 & 0.0131 & 0.0117 \\
 \hline
\end{tabular}
\tablefoot{The spectra are not extended with Kurucz models. The \textit{nGood} column lists the number of objects for which residuals can be calculated: a few objects do not have observed magnitudes in the \textit{u} and \textit{z} bands, and for this reason, \textit{nGood} is smaller for these bands.}
\end{table}

\begin{table}
\caption{Same as Tab.~\ref{tab:stat_noExt}, but for the spectra extended with cutoff values [350;1020]. }\label{tab:stat_350_1020}
\centering
\begin{tabular}{ |c||c|c|c|c|c| } 
 \hline
Band & nGood & $med^{synth}_{resid}$ & $med^{corr}_{resid}$ & $std^{synth}_{resid}$ & $std^{corr}_{resid}$ \\
 \hline\hline 
u & 137938 & 0.0203 & 0.0041 & 0.1144 & 0.1062 \\
g & 138770 & -0.0553 & -0.0002 & 0.0230 & 0.0119 \\
z & 138767 & -0.0279 & 0.0001 & 0.0126 & 0.0114 \\
 \hline
\end{tabular}
\tablefoot{The statistics for the \texttt{ri} bands are the same as for the non-extended spectra since extension with Kurucz affects only the tails of the spectrum.}
\end{table}

\begin{table}
\caption{Same as Tab.~\ref{tab:stat_noExt} and \ref{tab:stat_350_1020}, but for the spectra extended with cutoff values [400;1020].}\label{tab:stat_400_1020}
\centering
\begin{tabular}{ |c||c|c|c|c|c| } 
 \hline
Band & nGood & $med^{synth}_{resid}$ & $med^{corr}_{resid}$ & $std^{synth}_{resid}$ & $std^{corr}_{resid}$ \\
 \hline\hline 
u & 138761 & 0.0227 & 0.0020 & 0.1056 & 0.0707 \\
 \hline
\end{tabular}
\tablefoot{The statistics for the \texttt{griz} bands are the same as for the [350;1020] extended spectra.}
\end{table}

\begin{table}
\caption{Same as Tab.~\ref{tab:stat_350_1020}, but for the spectra corrected with DES magnitudes. Extension [350;1020].}
\label{tab:stat_350_1020_DES}
\centering
\begin{tabular}{ |c||c|c|c|c|c| } 
 \hline
Band & nGood & $med^{synth}_{resid}$ & $med^{corr}_{resid}$ & $std^{synth}_{resid}$ & $std^{corr}_{resid}$ \\
 \hline\hline 
g & 76917 & -0.0212 & -0.0000 & 0.0205 & 0.0103 \\
r & 76551 & -0.0027 & 0.0000 & 0.0086 & 0.0063 \\
i & 76121 & 0.0009 & 0.0002 & 0.0072 & 0.0057 \\
z & 76529 & 0.0049 & -0.0038 & 0.0111 & 0.0093 \\
y & 77045 & 0.0351 & 0.0000 & 0.0263 & 0.0201 \\
 \hline
\end{tabular}
\end{table}

\subsection{Comparison with the DES DR2 photometry and DES-based correction}\label{validationDES}

\begin{figure*}
\sidecaption
  \includegraphics[width=12cm]{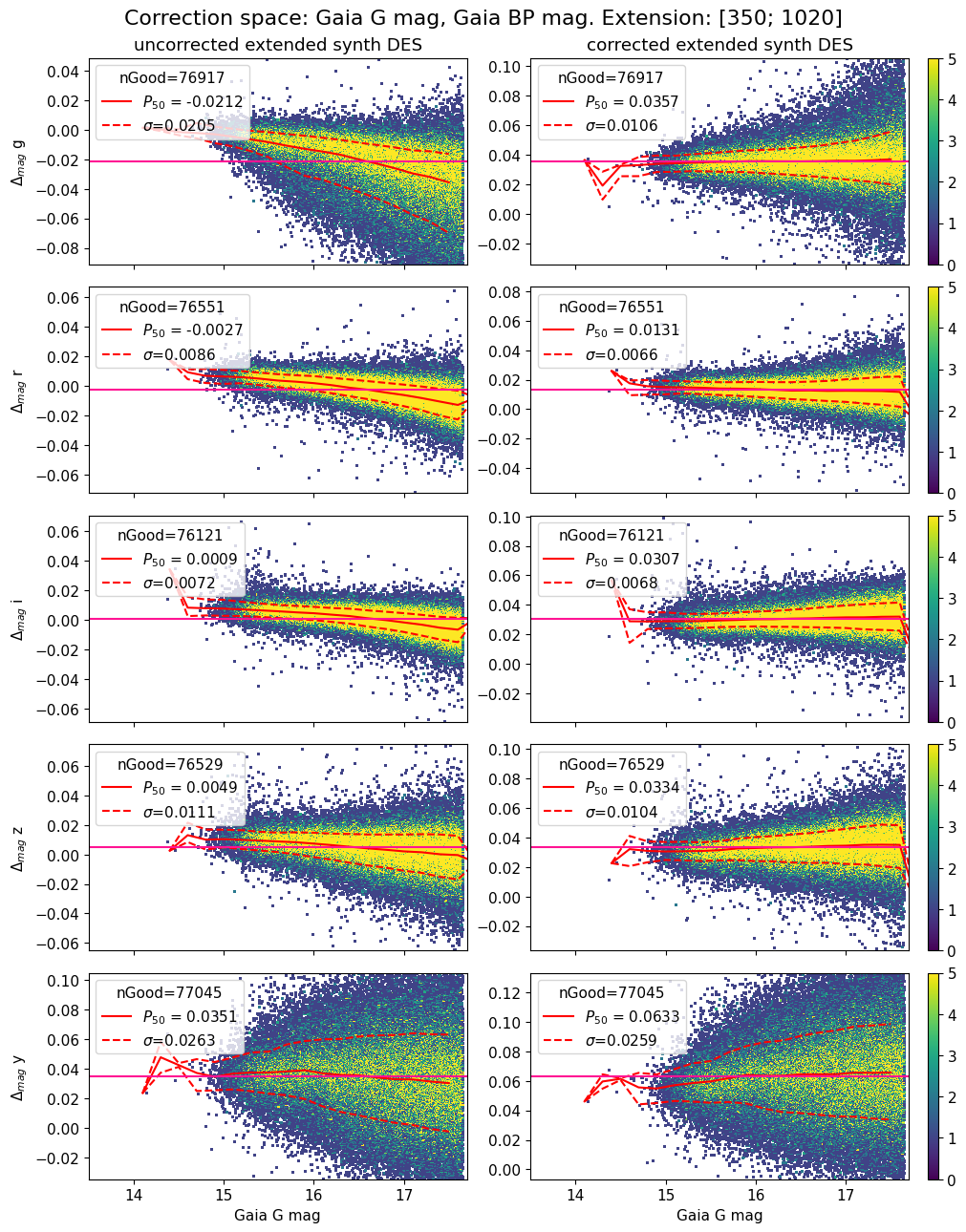}
     \caption{Magnitude-residual plots for the DES magnitudes for [350;1020] extended spectra. Left column: \texttt{rubin\_sim} synthetic DES magnitudes with Kurucz extension but without any correction. Right column: \texttt{rubin\_sim} synthetic DES magnitudes with Kurucz extension and after correction, described in Sect.~\ref{specCorr}. Please note that the horizontal pink line marks the median residual value, not 0, for better visibility of the magnitude-dependent term.}\label{fig:350_mag_des_medshift}
\end{figure*}

\begin{figure*}
\sidecaption
  \includegraphics[width=12cm]{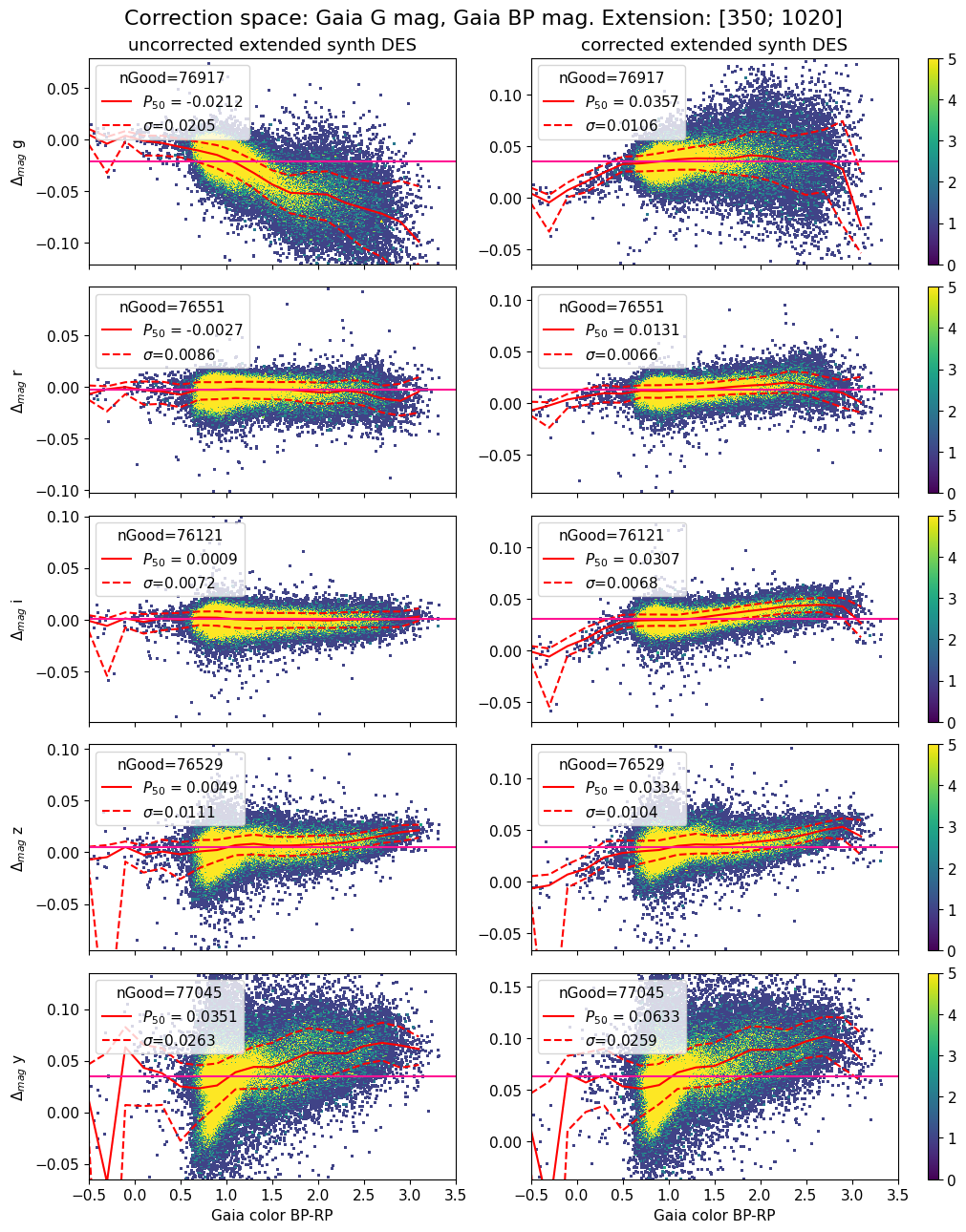}
     \caption{Colour-residual plots for the DES magnitudes for [350;1020] extended spectra. Left column: \texttt{rubin\_sim} synthetic DES magnitudes with Kurucz extension but without any correction. Right column: \texttt{rubin\_sim} synthetic DES magnitudes with Kurucz extension and after correction, described in Sect.~\ref{specCorr}. Please note that the horizontal pink line marks the median residual value, not 0, for better visibility of the colour-dependent term.}\label{fig:350_color_des_medshift}
\end{figure*}

\begin{figure*}
\sidecaption
  \includegraphics[width=12cm]{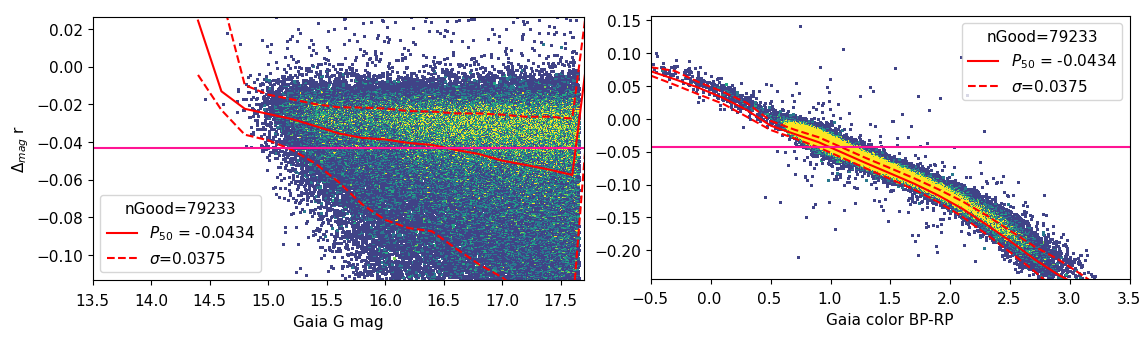}
     \caption{Magnitude- and colour-residual plots for the residuals between SDSS and DES observed magnitudes in the \textit{r} band. Please note that the horizontal pink line marks the median residual value, not 0, for better visibility of the magnitude- and colour-dependent terms. Other bands demonstrate similar behaviour.}\label{fig:DES_SDSS_resids}
\end{figure*}

As an additional validation, we produce synthetic magnitudes in a photometric system that was not used in the correction procedure, the \texttt{grizY} DES photometry. We calculate the synthetic magnitudes for the uncorrected and corrected Gaia XP spectra for the DES sample described in Sect.~\ref{DES} and compare them with the observed magnitudes (see Fig. \ref{fig:350_mag_des_medshift}, \ref{fig:350_color_des_medshift}). The improvement for these magnitudes is similar to that for SDSS: the scatter of residuals decreases by 10-20\% for the \textit{riz} bands and by the factor of 2 for the \textit{g} band. Magnitude-dependent terms are strongly reduced, with a slight overcorrection at the faint end. For colour-dependent terms, the overcorrection is more prominent. The median values of the residuals are also increased. Both overcorrection and increased median of residuals appear because observed SDSS and DES photometry are not perfectly calibrated against each other. The residuals between these two datasets demonstrate strong magnitude- and colour-dependent terms (see Fig. \ref{fig:DES_SDSS_resids}), so as we correct Gaia XP spectra to be in best agreement with the S82 observed magnitudes, we introduce stronger difference with the DES observations. 

Considering that Gaia XP spectra and observed DES photometry are in better agreement than Gaia XP and S82 photometry, we also perform Gaia XP correction using DES magnitudes to generate synthetic LSST photometry. The synthetic DES photometry before and after the correction shows similar behaviour as in the case of S82 correction (see Tab.~\ref{tab:stat_350_1020_DES}).

\subsection{Synthetic LSST magnitudes and comparison with Data Preview 1}\label{dp1_comparison}

\begin{table}
\caption{Statistics for the LSST synthetic magnitudes, compared with DP1 observations. }\label{tab:stat_noExt_LSST_s82}
\centering
\begin{tabular}{ |c||c|c|c|c|c| } 
 \hline
Band & nGood & $med^{synth}_{resid}$ & $med^{corr}_{resid}$ & $std^{synth}_{resid}$ & $std^{corr}_{resid}$ \\
 \hline\hline 
u & 2422 & -0.0670 & -0.0448 & 0.1125 & 0.1235 \\
g & 2440 & -0.0288 & 0.0229 & 0.0215 & 0.0103 \\
r & 2440 & -0.0017 & 0.0129 & 0.0084 & 0.0076 \\
i & 2440 & -0.0060 & 0.0238 & 0.0101 & 0.0084 \\
z & 2440 & 0.0043 & 0.0399 & 0.0089 & 0.0088 \\
y & 2440 & -0.0109 & 0.0239 & 0.0240 & 0.0233 \\
 \hline
\end{tabular}
\tablefoot{The spectra are corrected with S82 sample and have no extension.}
\end{table}

\begin{table}
\caption{Same as Tab.~\ref{tab:stat_noExt_LSST_s82}, but using [350;1020] extension.}\label{tab:stat_350_1020_LSST_s82}
\centering
\begin{tabular}{ |c||c|c|c|c|c| } 
 \hline
Band & nGood & $med^{synth}_{resid}$ & $med^{corr}_{resid}$ & $std^{synth}_{resid}$ & $std^{corr}_{resid}$ \\
 \hline\hline 
u & 2377 & -0.0719 & -0.0624 & 0.0978 & 0.1008 \\
g & 2387 & -0.0286 & 0.0229 & 0.0209 & 0.0101 \\
y & 2387 & 0.0119 & 0.0424 & 0.0202 & 0.0196 \\
 \hline
\end{tabular}
\tablefoot{\textit{riz} bands are the same as in Tab.~\ref{tab:stat_noExt_LSST_s82}.}
\end{table}

\begin{table}
\caption{Same as Tab.~\ref{tab:stat_noExt_LSST_s82}, but using [400;1020] extension.}\label{tab:stat_400_1020_LSST_s82}
\centering
\begin{tabular}{ |c||c|c|c|c|c| } 
 \hline
Band & nGood & $med^{synth}_{resid}$ & $med^{corr}_{resid}$ & $std^{synth}_{resid}$ & $std^{corr}_{resid}$ \\
 \hline\hline 
u & 2387 & -0.0338 & -0.0574 & 0.0809 & 0.0617 \\
g & 2387 & -0.0284 & 0.0230 & 0.0207 & 0.0101 \\
 \hline
\end{tabular}
\tablefoot{\textit{rizy} bands are the same as in Tab.~\ref{tab:stat_350_1020_LSST_s82}.}
\end{table}

\begin{table}
\caption{Same as Tab.~\ref{tab:stat_noExt_LSST_s82}-\ref{tab:stat_350_1020_LSST_s82}, but for spectra corrected with DES sample and having no extension.}\label{tab:stat_noExt_LSST_DES}
\centering
\begin{tabular}{ |c||c|c|c|c|c| } 
 \hline
Band & nGood & $med^{synth}_{resid}$ & $med^{corr}_{resid}$ & $std^{synth}_{resid}$ & $std^{corr}_{resid}$ \\
 \hline\hline 
u & 2422 & -0.0670 & -0.0531 & 0.1125 & 0.1117 \\
g & 2440 & -0.0288 & -0.0107 & 0.0215 & 0.0113 \\
r & 2440 & -0.0017 & 0.0041 & 0.0084 & 0.0079 \\
i & 2440 & -0.0060 & -0.0046 & 0.0101 & 0.0097 \\
z & 2440 & 0.0043 & 0.0044 & 0.0089 & 0.0089 \\
y & 2440 & -0.0109 & -0.0146 & 0.0240 & 0.0230 \\
 \hline
\end{tabular}
\end{table}

\begin{table}
\caption{Same as Tab.~\ref{tab:stat_noExt_LSST_DES}, but with [350;1020] extension.}\label{tab:stat_350_1020_LSST_DES}
\centering
\begin{tabular}{ |c||c|c|c|c|c| } 
 \hline
Band & nGood & $med^{synth}_{resid}$ & $med^{corr}_{resid}$ & $std^{synth}_{resid}$ & $std^{corr}_{resid}$ \\
 \hline\hline 
u & 2377 & -0.0719 & -0.0572 & 0.0978 & 0.0936 \\
g & 2387 & -0.0286 & -0.0106 & 0.0209 & 0.0110 \\
r & 2387 & -0.0016 & 0.0040 & 0.0084 & 0.0077 \\
i & 2387 & -0.0058 & -0.0045 & 0.0100 & 0.0095 \\
z & 2387 & 0.0044 & 0.0098 & 0.0089 & 0.0088 \\
y & 2387 & 0.0119 & -0.0097 & 0.0202 & 0.0179 \\
 \hline
\end{tabular}
\end{table}

\begin{table}
\caption{Same as Tab.~\ref{tab:stat_350_1020_LSST_DES}, but with [400;1020] extension.}\label{tab:stat_400_1020_LSST_DES}
\centering
\begin{tabular}{ |c||c|c|c|c|c| } 
 \hline
Band & nGood & $med^{synth}_{resid}$ & $med^{corr}_{resid}$ & $std^{synth}_{resid}$ & $std^{corr}_{resid}$ \\
 \hline\hline 
u & 2387 & -0.0338 & -0.0178 & 0.0809 & 0.0746 \\
g & 2387 & -0.0284 & -0.0107 & 0.0207 & 0.0110 \\
 \hline
\end{tabular}
\tablefoot{\textit{rizy} bands are the same as in Tab.~\ref{tab:stat_350_1020_LSST_DES}.}
\end{table}

With the corrected spectra, we produce synthetic LSST magnitudes and compare them with the DP1 sample. Both S82- and DES-based corrections are beneficial for the synthetic magnitudes, but in different ways (see Tab.~\ref{tab:stat_noExt_LSST_s82}-\ref{tab:stat_400_1020_LSST_DES}). 

As with S82 and DES magnitudes, even extended spectra without correction show lower scatter in \textit{ug} and \textit{y} bands. The magnitude- and colour-dependent terms are present for all bands, however, the correction procedure strongly reduces them. The main and completely expected difference between S82- and DES-based correction is that S82-based correction works better for reducing the scatter and colour-dependent terms in all bands, except \textit{y}. This difference is explained by the fact that SDSS and LSST filters are more similar than DES and LSST. However, DES-based correction is needed for the \textit{y} band, for which SDSS simply does not provide the correction. It is also worth noting that S82-based correction introduces large median residuals, similarly to how it happens with synthetic DES photometry obtained from the spectra corrected with S82 observations. In other words, there is a zero-point offset between S82 and LSST magnitudes.

In principle, it is possible to combine the corrected magnitudes obtained with different samples, since the correction is effectively done by piecewise multiplication of parts of the spectra with independently derived correction coefficients. In this case we would recommend using \textit{ugri} photometry corrected with S82 sample and \textit{zy} magnitudes corrected with the DES sample, with subtraction of the corresponding median residuals.
 
\subsection{Run catalogue}\label{catDescription}

The full version of the catalogue includes synthetic SDSS, DES and LSST magnitudes for the uncorrected, extended without correction, and extended and corrected Gaia XP spectra, median and standard deviation of the magnitude residuals within the magnitude-magnitude cell to which the object belongs (where the magnitude residuals are calculated for the objects that were used for deriving correction coefficients), and the number of objects from the S82 or DES catalogue used to calculate these values.
The median and standard deviation of residuals are reported as a proxy for magnitude errors. 
The Gaia XP spectra are provided with the full covariance matrix, so it is theoretically possible to calculate magnitude errors for each band. However, our correction procedure introduces uncertainties that are difficult to quantify, e.g. those related to the selection and interpolation of the stellar atmosphere models or those related to the corrections on the mag-mag grid itself. It is hard to determine a straightforward estimate of how these operations affect flux errors. While median and standard deviation of residuals depend on the resolution of the magnitude-magnitude grid used for the correction, the upside of this approach is that it is strictly empirical and adequately represents the real differences between the synthetic magnitudes and observed ones.

The counts of objects with valid magnitudes vary from case to case and from band to band. For completeness, we keep all the objects that have Gaia BP and RP spectra in the catalogue. About 86\% of them have extended and corrected magnitudes. There are two main reasons why an object misses extended and/or corrected magnitudes. One reason is the absence of astrophysical parameters in the Gaia Archive, needed for determination of the best-fit Kurucz model. Another is the situation when the object falls into a cell of the magnitude-magnitude grid that has less than five objects from S82 or DES catalogue. In these cases, the correction coefficient equals 1, so the corrected magnitude is the same as uncorrected.

As LSST Data Preview 2 and in particular Data Release 1 become available, we plan to use them for deriving more precise corrections for the Gaia XP spectra.

\section{Conclusions}\label{Conclusions}

In this paper, we aimed to create a synthetic LSST photometry catalogue using low-resolution Gaia XP spectra. Since these spectra are known to have certain magnitude- and colour-dependent biases, we used the Stripe 82 catalogue as a ground truth for correcting them. Another known issue with the Gaia XP spectra, so-called wiggles, originating from the correlations of the coefficients of the basis functions used for storing continuous spectra, were compensated by replacing the problematic parts of the spectra with the best-fit and interpolated Kurucz stellar atmosphere models.

After the extension and correction, the residuals between the synthetic SDSS and observed Stripe 82 magnitudes in the best case have a median absolute offset of less than $0.0002$ mag and the scatter (inter-quantile range) of $\lesssim 0.01$ for the \texttt{griz} bands. For the \textit{u}-band the median offset is $\sim0.002$ mag and the scatter is $\le0.1$. A validation performed with DES DR2 and LSST DP1 photometric data shows that the correction also improves the magnitude- and colour-dependent bias for this photometry.

We present the synthetic magnitude catalogue that contains SDSS, DES, and LSST magnitudes, together with statistical quality indicators for these magnitudes. The code for reproducing the analysis is available in the GitHub repository \url{https://github.com/ShrRa/LSST_Gaia_XPy_public}. 
Assuming that the S82 photometry is the best approximation of the future LSST observations we currently have, the produced synthetic photometric catalogue is suitable for use alongside with the early LSST photometry as it is. After the first LSST Data Release it will be possible to perform a better correction and validation of the synthetic magnitudes using LSST data themselves. However, it is worth noting that LSST and Gaia XP datasets will have a relatively small overlap in terms of magnitude ranges. Provided that the S82 photometry does not show any major biases in comparison with the LSST data, it will be beneficial to correct Gaia XP spectra using both datasets.

\section{Data availability}
The full version of the catalogue is available in  \href{https://data.fulir.irb.hr/en/object/irb:909}{the Ruđer Bošković Institute FULIR Repository}. The S82, DES, and LSST DP 1 samples are available in \href{https://zenodo.org/records/21417949}{Zenodo data repository} \citep{Razim2026-cd}. 

\begin{acknowledgements}

Funding: This work was financed within the Tenure Track Pilot Programme of the Croatian Science Foundation and the Ecole Polytechnique Fédérale de Lausanne and the Project TTP-2018-07-1171 Mining the variable sky, with the funds of the Croatian-Swiss Research Programme and is supported by the Croatian Science Foundation under the project number IP-2025-02-1942.
This publication is co-funded by 
the European Union’s Horizon Europe research and innovation program under the
Marie Sklodowska-Curie COFUND Postdoctoral Programme grant agreement
No.101081355-SMASH and by the Republic of Slovenia and the European
Union from the European Regional Development Fund.

Gaia: This work has made use of data from the European Space Agency (ESA) mission
\textit{Gaia} (\url{https://www.cosmos.esa.int/gaia}), processed by the \textit{Gaia}
Data Processing and Analysis Consortium (DPAC,
\url{https://www.cosmos.esa.int/web/gaia/dpac/consortium}). Funding for the DPAC
has been provided by national institutions, in particular the institutions
participating in the \textit{Gaia} Multilateral Agreement.

SDSS-IV, DR15: Funding for the Sloan Digital Sky Survey IV has been provided by the Alfred P. Sloan Foundation, the U.S. Department of Energy Office of Science, and the Participating Institutions. SDSS acknowledges support and resources from the Center for High-Performance Computing at the University of Utah. The SDSS web site is www.sdss4.org.

SDSS is managed by the Astrophysical Research Consortium for the Participating Institutions of the SDSS Collaboration including the Brazilian Participation Group, the Carnegie Institution for Science, Carnegie Mellon University, Center for Astrophysics | Harvard \& Smithsonian (CfA), the Chilean Participation Group, the French Participation Group, Instituto de Astrofísica de Canarias, The Johns Hopkins University, Kavli Institute for the Physics and Mathematics of the Universe (IPMU) / University of Tokyo, the Korean Participation Group, Lawrence Berkeley National Laboratory, Leibniz Institut für Astrophysik Potsdam (AIP), Max-Planck-Institut für Astronomie (MPIA Heidelberg), Max-Planck-Institut für Astrophysik (MPA Garching), Max-Planck-Institut für Extraterrestrische Physik (MPE), National Astronomical Observatories of China, New Mexico State University, New York University, University of Notre Dame, Observatório Nacional / MCTI, The Ohio State University, Pennsylvania State University, Shanghai Astronomical Observatory, United Kingdom Participation Group, Universidad Nacional Autónoma de México, University of Arizona, University of Colorado Boulder, University of Oxford, University of Portsmouth, University of Utah, University of Virginia, University of Washington, University of Wisconsin, Vanderbilt University, and Yale University.

DES DR2: This project used public archival data from the Dark
Energy Survey (DES). Funding for the DES Projects has been provided by the U.S. Department of Energy, the U.S. National Science
Foundation, the Ministry of Science and Education of Spain, the
Science and Technology FacilitiesCouncil of the United Kingdom,
the Higher Education Funding Council for England, the National
Center for Supercomputing Applications at the University of Illinois
at Urbana-Champaign, the Kavli Institute of Cosmological Physics
at the University of Chicago, the Center for Cosmology and AstroParticle Physics at the Ohio State University, the Mitchell Institute
for Fundamental Physics and Astronomy at Texas A\&M University, Financiadora de Estudos e Projetos, Fundação Carlos Chagas
Filho de Amparo à Pesquisa do Estado do Rio de Janeiro, Conselho
Nacional de Desenvolvimento Científico e Tecnológico and the Ministério da Ciência, Tecnologia e Inovação, the Deutsche Forschungsgemeinschaft, and the Collaborating Institutions in the Dark Energy
Survey. The Collaborating Institutions are Argonne National Laboratory, the University of California at Santa Cruz, the University of
Cambridge, Centro de Investigaciones Energéticas, Medioambientales y Tecnológicas-Madrid, the University of Chicago, University
College London, the DES-Brazil Consortium, the University of Edinburgh, the Eidgenössische Technische Hochschule (ETH) Zürich,
Fermi National Accelerator Laboratory, the University of Illinois at
Urbana-Champaign, the Institut de Ciències de l’Espai (IEEC/CSIC),
the Institut de Física d’Altes Energies, Lawrence Berkeley National
Laboratory, the Ludwig-Maximilians Universität München and the
associated Excellence Cluster Universe, the University of Michigan, the National Optical Astronomy Observatory, the University
of Nottingham, The Ohio State University, the OzDES Membership Consortium, the University of Pennsylvania, the University of
Portsmouth, SLAC National Accelerator Laboratory, Stanford University, the University of Sussex, and Texas A\&M University. Based
in part on observations at Cerro Tololo Inter-American Observatory, National Optical Astronomy Observatory, which is operated by the
Association of Universities for Research in Astronomy (AURA) under a cooperative agreement with the National Science Foundation.

LSST: This material is based upon work supported in part by the National Science Foundation through Cooperative Agreements AST-1258333 and AST-2241526 and Cooperative Support Agreements AST-1202910 and 2211468 managed by the Association of Universities for Research in Astronomy (AURA), and the Department of Energy under Contract No. DE-AC02-76SF00515 with the SLAC National Accelerator Laboratory managed by Stanford University. Additional Rubin Observatory funding comes from private donations, grants to universities, and in-kind support from LSST-DA Institutional Members.

The authors are deeply grateful to the anonymous reviewer for a thorough analysis of the paper and useful suggestions on readability and structure.

Disclaimer: Co-funded by the European Union. Views and opinions
expressed are however those of the authors only and do not necessarily reflect
those of the European Union or European Research Exacutive Agency. Neither the
European Union nor the granting authority can be held responsible for them.

Generative AI usage disclosure: Claude Sonnet 4.6/ChatGPT-5.5 were used for grammar and style corrections.
Opus 4.7, 4.8, and Sonnet 4.6 were used for code vectorization, bug search, refactoring and documentation. The authors carefully reviewed suggested edits and are fully responsible for the content of the paper. 
OR acknowledges ChatGPT-4o 'for dragging this paper across the finish line with motivational bullying and structurally sound sarcasm' and Sonnet 4.6/Opus 4.6 for organisational and emotional support. 

Software citations: 
Numpy \citep{numpy}, 
Pandas \citep{pandas}, 
Matplotlib \citep{matplotlib}, 
Scipy \citep{scipy}, 
Astropy \citep{astropy}, 
LSDB \citep{lsdb},
rubin\_sim \citep{rubinsim},
GaiaXPy \citep{gaiaxpy}, 
extinction \citep{BarbaryExtinct}.
\end{acknowledgements}

\bibliographystyle{aa}
\bibliography{references}

\newpage
\begin{appendix}
\section{Quality cuts used to obtain the correction and validation data sample}\label{appendixFilters}

\begin{itemize}
    \item XP spectra quality: we use only objects with more than 15 observations, coming from more than 10 visibility periods\footnote{A visibility period is a group of observations separated from other groups by a gap of at least 4 days. Full description is available at \href{https://gea.esac.esa.int/archive/documentation/GDR3/Gaia_archive/chap_datamodel/sec_dm_main_source_catalogue/ssec_dm_gaia_source.html}{Gaia documentation}.} and having fewer than 10\% of contaminated or blended observations. 

    The ADQL query for these criteria:
\begin{verbatim}
select source_id, ra, dec from 
gaiadr3.gaia_source as gaia 
where ((gaia.ra between 0 and 60 
and gaia.dec between -1.26 and 1.26)
or (gaia.ra between 308 and 360 
and gaia.dec between -1.26 and 1.26)) 
and gaia.has_xp_continuous='t' and
gaia.phot_rp_n_obs > 15 
and gaia.phot_bp_n_obs > 15 and
gaia.phot_rp_n_contaminated_transits/
gaia.phot_rp_n_obs <0.1
and gaia.phot_bp_n_contaminated_transits/
gaia.phot_bp_n_obs <0.1
and gaia.phot_rp_n_blended_transits/
gaia.phot_rp_n_obs <0.1
and gaia.phot_bp_n_blended_transits/
gaia.phot_bp_n_obs <0.1
and gaia.visibility_periods_used>10
        \end{verbatim}
    
    These selection criteria return 269861 objects with Gaia XP spectra in the $-52^\circ \le RA \le 60^\circ$, $-1.266^\circ \le DEC \le 1.266^\circ$ region.
    
    \item SDSS Stripe 82 photometry quality: the selected sources have magnitudes calculated from more than 4 observations in each band and a high signal-to-noise ratio in the \textit{gri} bands. 

    $\{u,g,r,i,z\}Nobs > 4; \{g, r, i\}msig \cdot \sqrt{\{g,r,i\}Nobs} < 0.03$
    
    These criteria return 434562 objects from the Stripe 82 catalogue out of the original $\sim1$ million. 
    
    \item The uniqueness of the SDSS S82 - Gaia cross-match, verified using the \texttt{gaiadr3.sdssdr13\_best\_neighbour} table in the \href{https://gea.esac.esa.int/archive/}{Gaia Archive}. From the sample obtained in the previous step, only those sources for which a single Gaia source is matched to a single SDSS S82 source are used. 
    
\begin{verbatim}
    select ... from 
    gaiadr3.sdssdr13_best_neighbour where
    jointab.number_of_neighbours = 1 and 
    jointab.number_of_mates = 0
\end{verbatim}
    
    These criteria return a total of 433095 sources. 
    
    \item Finally, from the catalogue from the previous step, we select objects that are also present in the table from the first step. This way we obtain objects with high-quality SDSS photometry, high-quality XP spectra, and a secure cross-match. After this step, we have 146550 objects.
\end{itemize}

These criteria apply only to the sample used for deriving the correction coefficients. The full synthetic magnitude catalogue contains all the sources with continuous Gaia XP spectra available in the Gaia Archive, regardless of their quality parameters.

For the LSST DP1 Object catalogue, we use the LSDB \cite{lsdb} request equivalent to the query:
\begin{verbatim}
select objectId,coord_dec,coord_decErr,coord_ra,
coord_raErr,shape_flag,z_extendedness,
z_extendedness_flag,u_extendedness,
u_extendedness_flag,
g_extendedness,g_extendedness_flag,
r_extendedness,r_extendedness_flag,i_extendedness,
i_extendedness_flag,y_extendedness,
y_extendedness_flag,u_psfMag,u_psfMagErr,g_psfMag,
g_psfMagErr,r_psfMag,r_psfMagErr,i_psfMag,
i_psfMagErr,z_psfMag,z_psfMagErr,y_psfMag,
y_psfMagErr from lsst.object where
shape_flag == 0 and u_extendedness <0.02 
and g_extendedness <0.02 and r_extendedness <0.02 
and i_extendedness <0.02 and 
z_extendedness <0.02 and y_extendedness <0.02 and 
u_extendedness_flag == 0 and 
g_extendedness_flag == 0 and 
r_extendedness_flag == 0 and 
i_extendedness_flag == 0 and 
z_extendedness_flag == 0 and 
y_extendedness_flag == 0
\end{verbatim}

Then we crossmatch the returned catalogue with Gaia DR3, selecting only one best neighbour in 1 arcsec radius. This crossmatch produces 8\,183 objects, out of which only 2\,440 have XP spectra.

\section{Systematic biases between synthetic and observed photometry}\label{appendixPlots}
\begin{figure*}
\sidecaption
  \includegraphics[width=12cm]{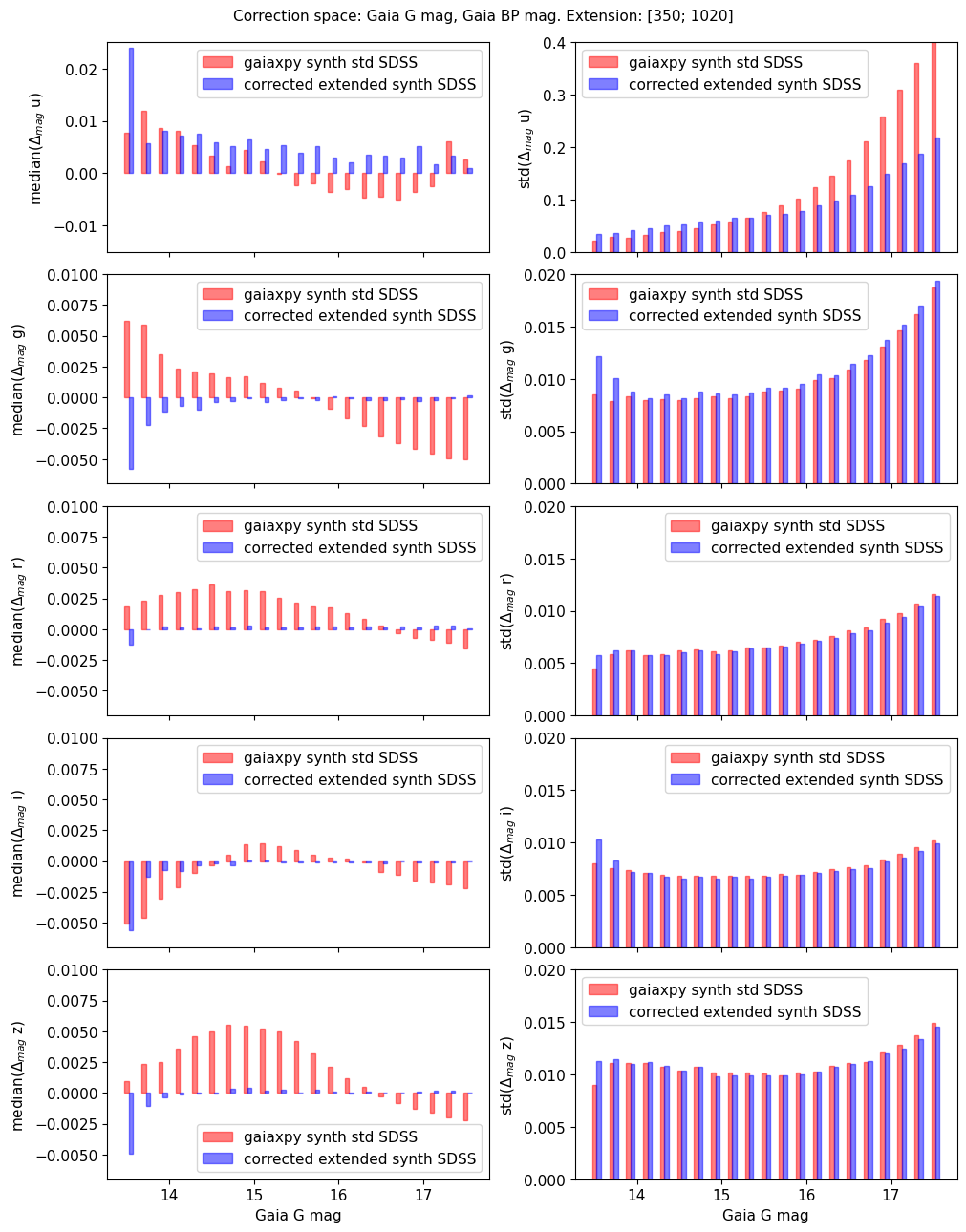}
     \caption{The statistics for the residuals between synthetic and observed S82 photometry in photometric bins. For \texttt{rubin\_sim} photometry, the spectra with [350;1020] extension and S82 correction were used. Large median residuals at the bright end are explained by a low number of objects available there for deriving the correction.}\label{fig:SDSS_binned_stats_full}
\end{figure*}

 \begin{figure*}
\sidecaption
  \includegraphics[width=12cm]{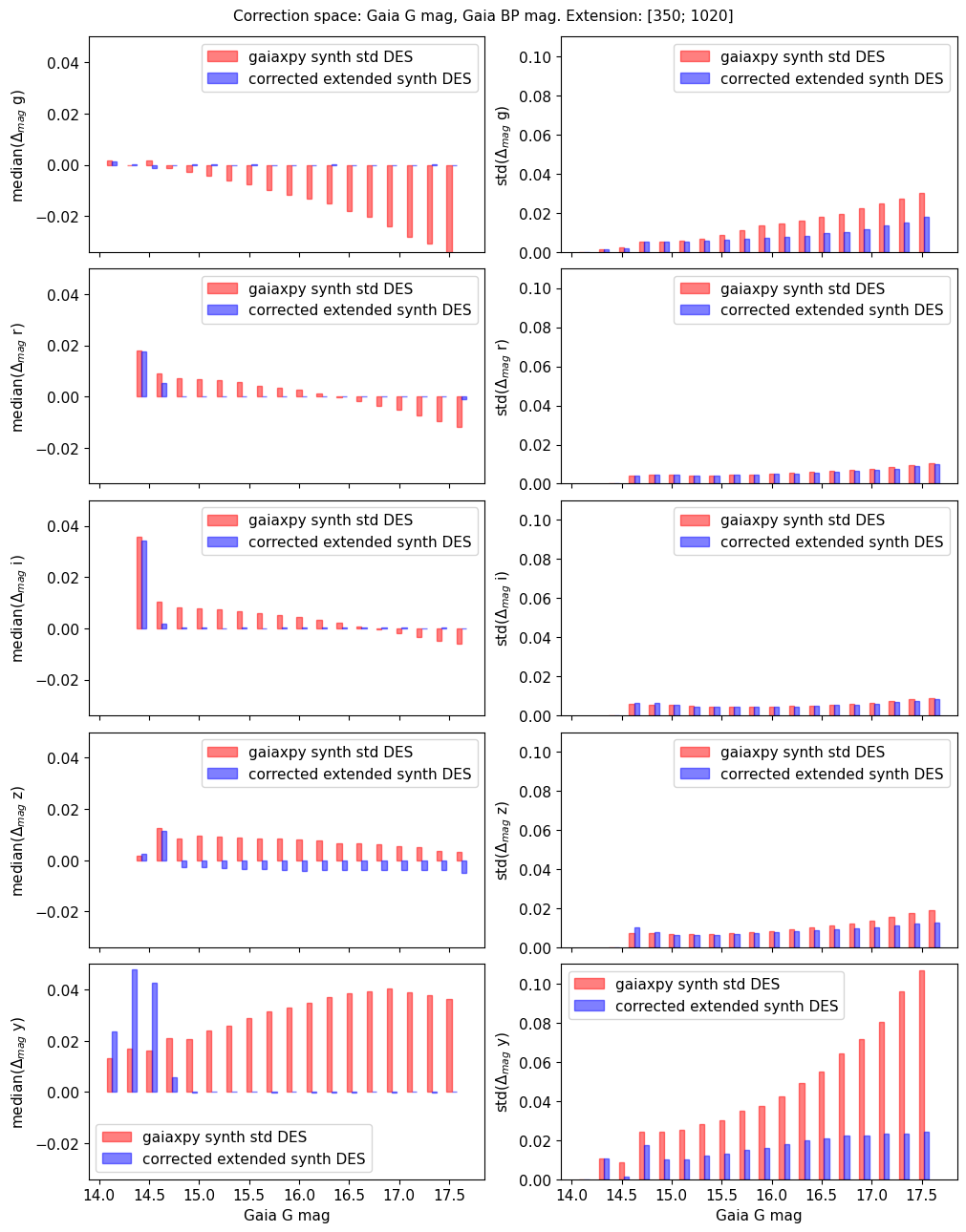}
     \caption{Residuals between synthetic and observed DES photometry. For \texttt{rubin\_sim} photometry, the spectra with [350;1020] extension and DES-based correction were used.}\label{fig:DES_binned_stats_full}
\end{figure*}

 \begin{figure*}
\sidecaption
  \includegraphics[width=12cm]{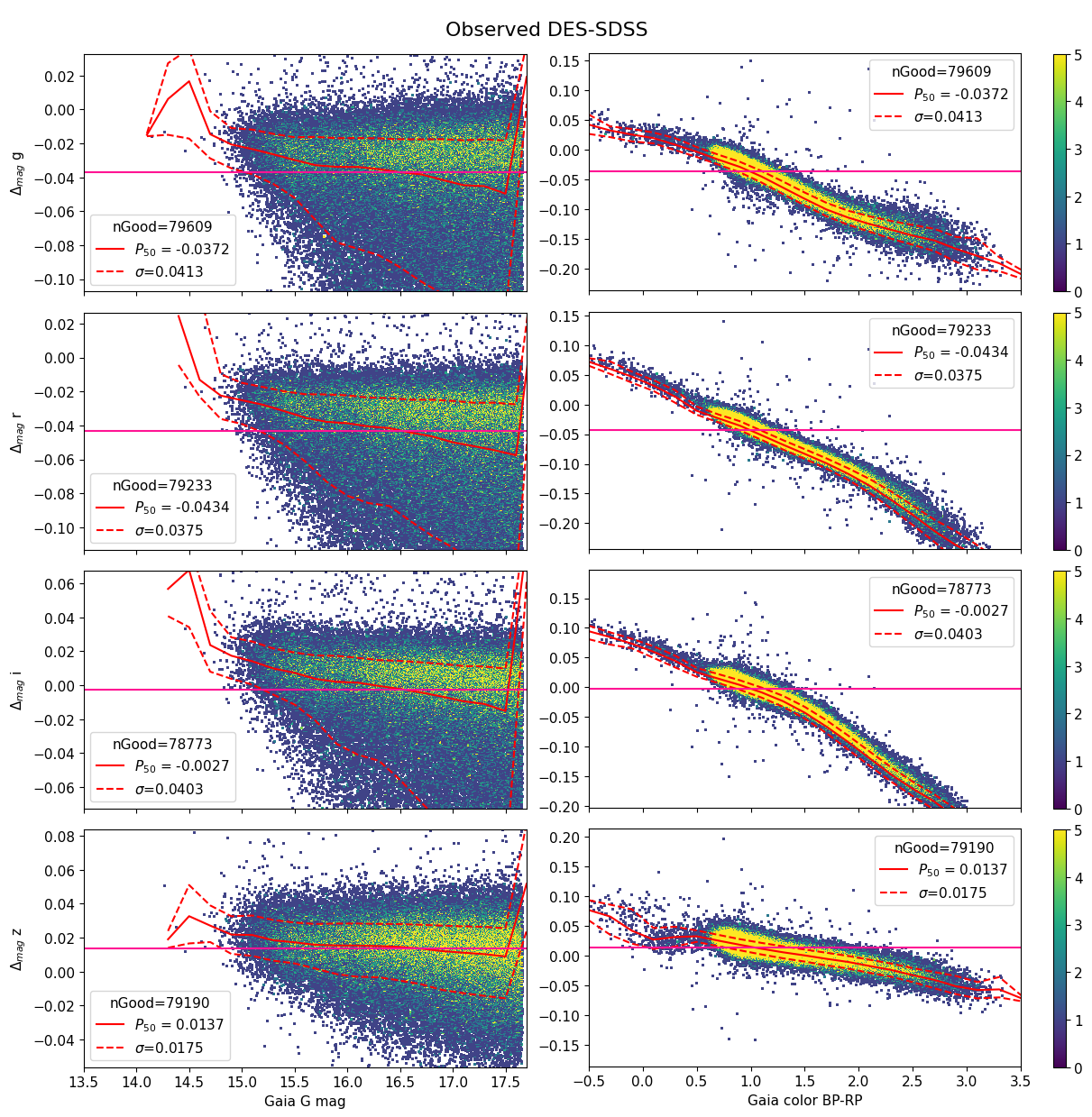}
     \caption{Magnitude- and colour-residual plots for the residuals between SDSS and DES observed magnitudes. Please note that the horizontal pink line marks the median residual value, not 0, for better visibility of the magnitude- and colour-dependent terms. These residual differences explain constant offsets and colour-dependent overcorrection for synthetic DES photometry when the correction is done with S82 observed photometry, and vice versa.}\label{fig:DES_SDSS_resids_both}
\end{figure*}

\begin{figure*}
\includegraphics[width=0.98\textwidth, center]{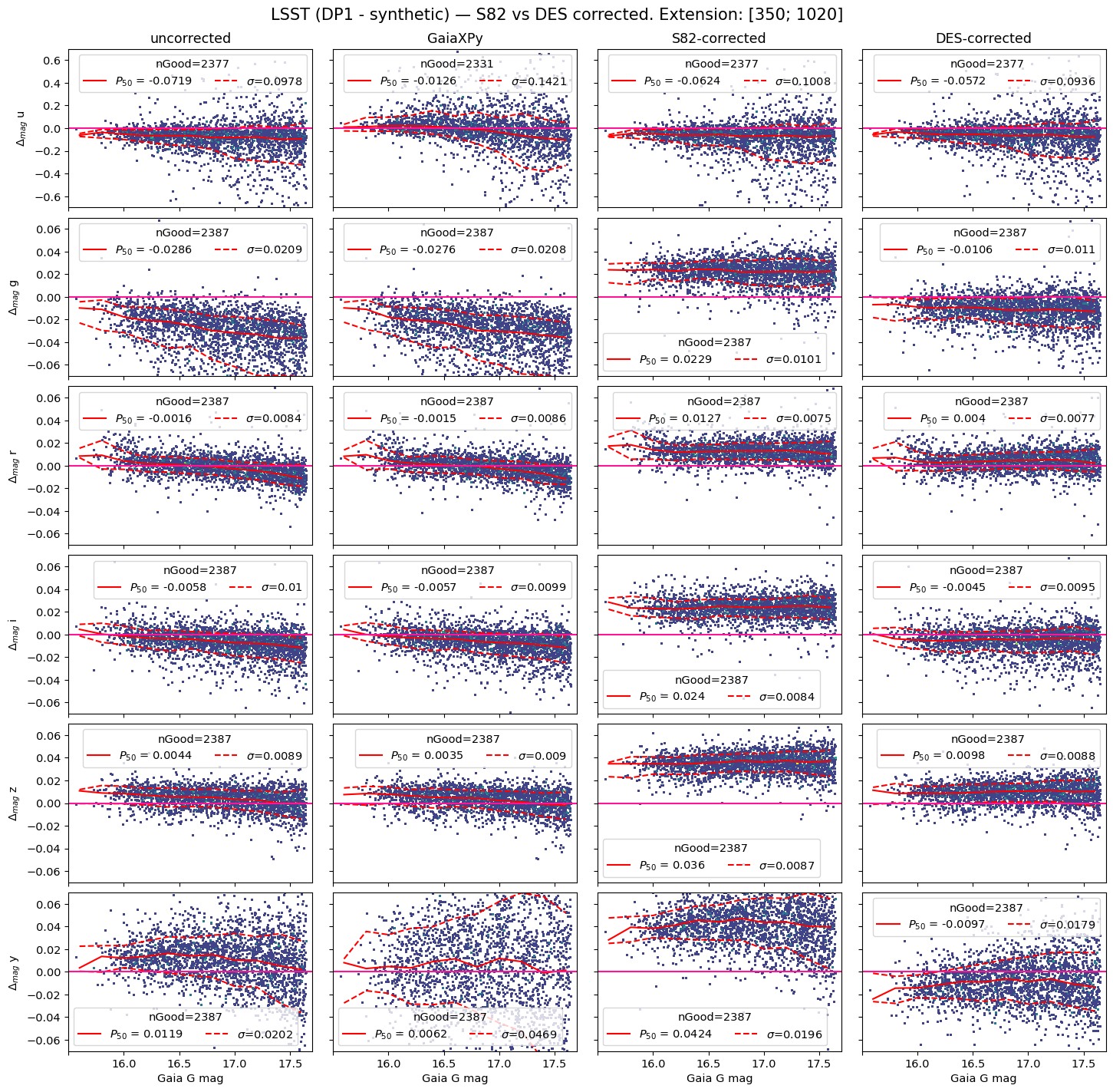}
    \caption{Magnitude-residual plots for the LSST magnitudes for [350;1020] extended spectra. First column: \texttt{rubin\_sim} synthetic LSST magnitudes with Kurucz extension but without any correction. 
    Second column: \texttt{GaiaXPy} synthetic LSST magnitudes. Third column: \texttt{rubin\_sim} magnitudes corrected using S82 sample. Fourth column: \texttt{rubin\_sim} magnitudes corrected using DES sample.}\label{fig:350_mag_lsst}
\end{figure*}

\begin{figure*}
\includegraphics[width=0.98\textwidth, center]{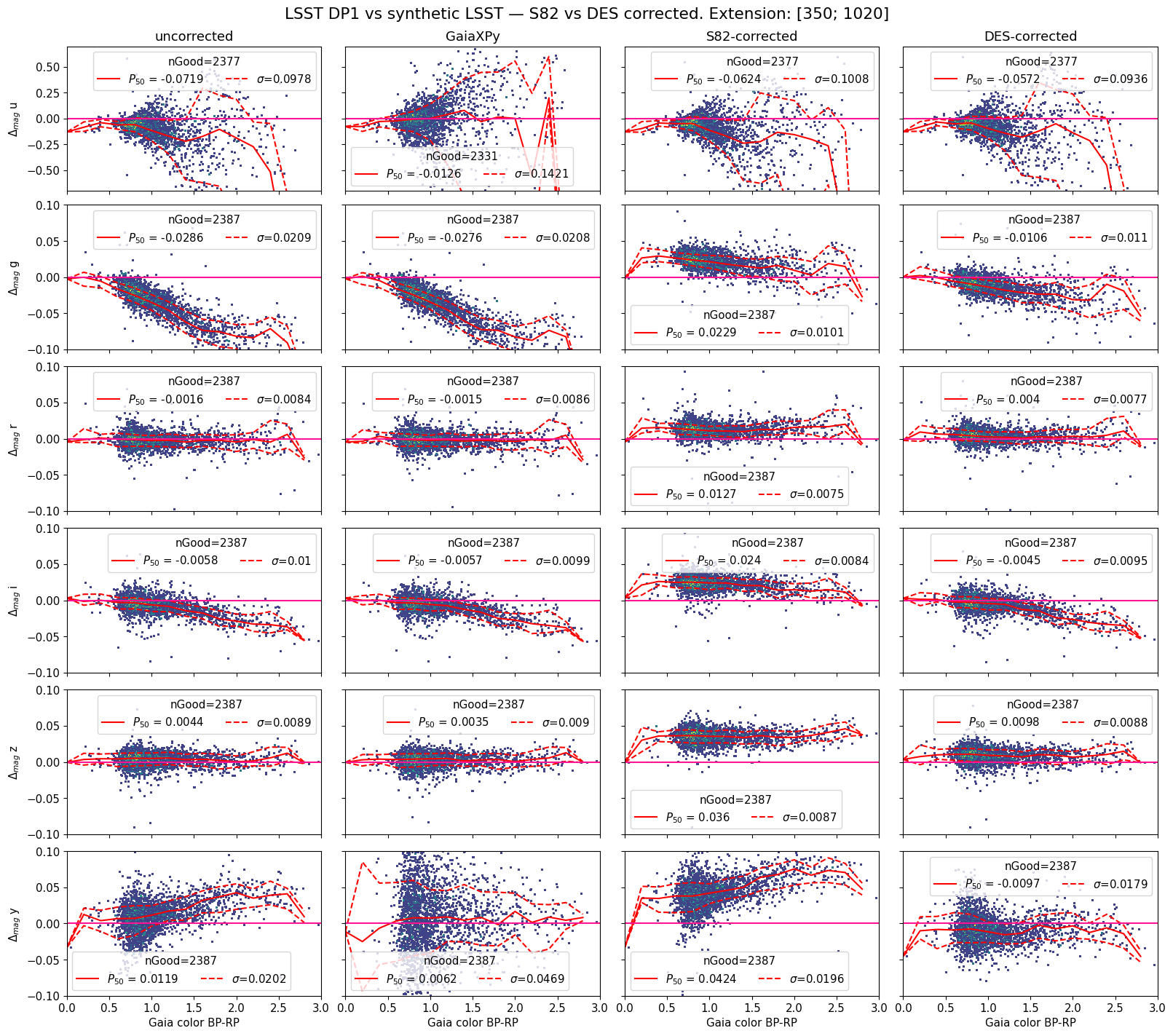}
    \caption{Colour-residual plots for the LSST magnitudes for [350;1020] extended spectra. First column: \texttt{rubin\_sim} synthetic LSST magnitudes with Kurucz extension but without any correction. 
    Second column: \texttt{GaiaXPy} synthetic LSST magnitudes. Third column: \texttt{rubin\_sim} magnitudes corrected using S82 sample. Fourth column: \texttt{rubin\_sim} magnitudes corrected using DES sample.}\label{fig:350_color_lsst}
\end{figure*}

\begin{figure*}
\includegraphics[width=0.98\textwidth, center]{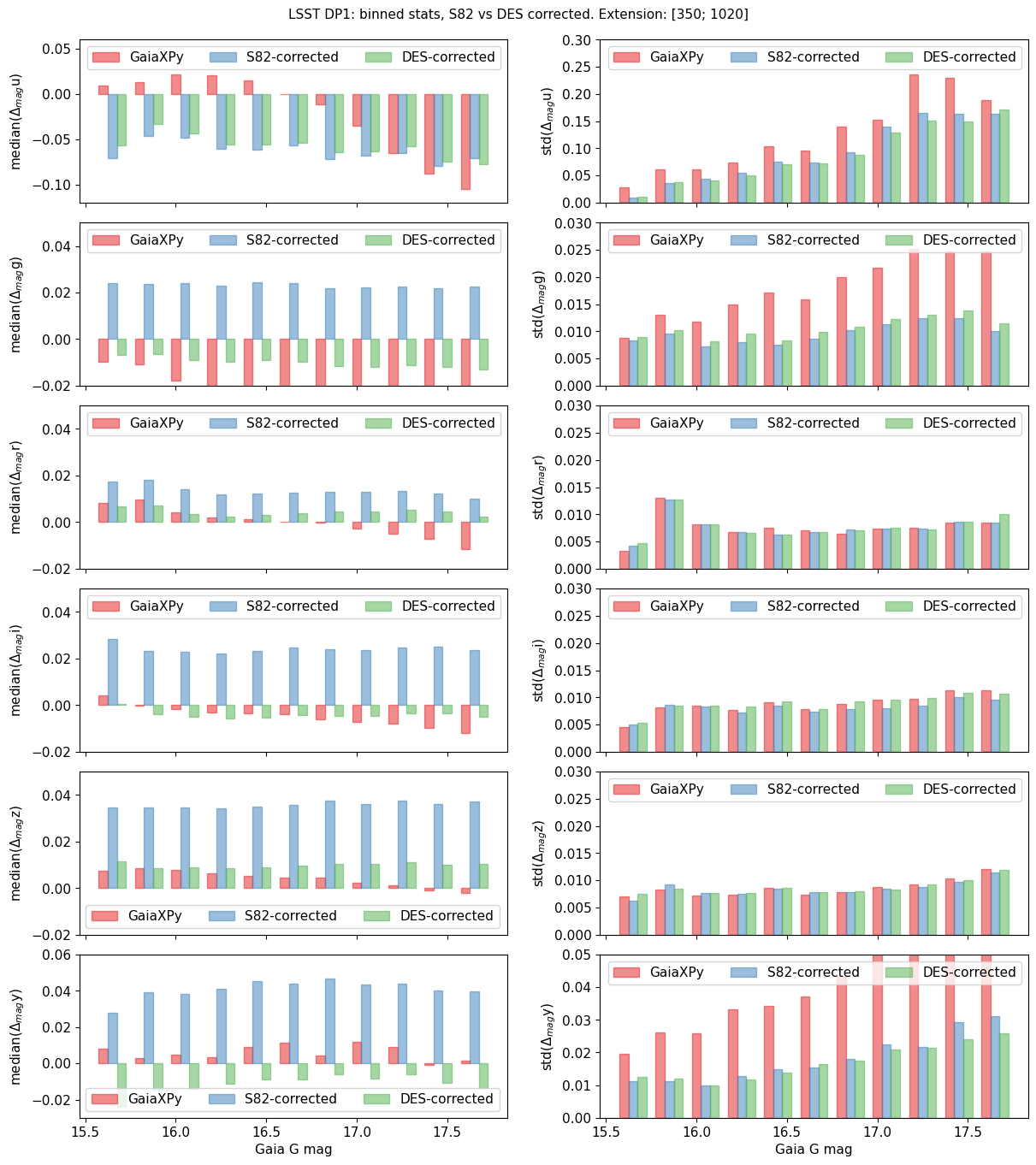}
    \caption{Statistics for residuals for the LSST magnitudes in magnitude bins for \texttt{GaiaXPy}, \texttt{rubin\_sim} corrected with S82, and \texttt{rubin\_sim} corrected with DES photometry. Extension [350;1020]}\label{fig:350_1020_LSST_binned}
\end{figure*}

\end{appendix}
\end{document}